\documentclass[12pt]{article}
\usepackage{amssymb}
\begin{document}

\makeatletter
\def\@maketitle{\newpage
 \null
 {\normalsize \tt \begin{flushright} 
  \begin{tabular}[t]{l} \@date  
  \end{tabular}
 \end{flushright}}
 \begin{center} 
 \vskip 2em
 {\LARGE \@title \par} \vskip 1.5em {\large \lineskip .5em
 \begin{tabular}[t]{c}\@author 
 \end{tabular}\par} 
 \end{center}
 \par
 \vskip 1.5em} 
\makeatother
\topmargin=-1cm
\oddsidemargin=1.5cm
\evensidemargin=-.0cm
\textwidth=15.5cm
\textheight=23cm
\setlength{\baselineskip}{16pt}
\title{Matrix Configurations for Spherical 4-branes and Non-commutative 
Structures on  $S^4$ }
\author{  Ryuichi~{\sc Nakayama}\thanks{ nakayama@particle.sci.hokudai.ac.jp}
and Yusuke~{\sc Shimono}\thanks{yshimono@particle.sci.hokudai.ac.jp}
       \\[1cm]
{\small
    Division of Physics, Graduate School of Science,} \\
{\small
           Hokkaido University, Sapporo 060-0810, Japan}
}
\date{
  EPHOU-03-008  \\
hep-th/0402010 \\ 
February  2004  
}
%
%
\maketitle

\begin{abstract} 
We present a Matrix theory action and Matrix configurations for spherical 
4-branes. The dimension of the representations is given by $N=2(2j+1)$  
$\ ( j=1/2,1,3/2,...)$. The algebra which defines these 
configurations is not invariant under $SO(5)$ rotations but under 
$SO(3) \otimes SO(2)$.
We also construct a non-commutative product $\star$ for field theories 
on $S^4$ in terms of that on $S^2$. An explicit formula of the non-commutative 
product which corresponds to the $N=4$ dim representation of the 
non-commutative $S^4$ algebra is worked out. Because we use $S^2 \otimes S^2$ 
parametrization of $S^4$, our $S^4$ is doubled and 
 the non-commutative product and functions on 
$S^4$ are indeterminate on a great circle ($S^1$) on $S^4$. We will however, 
show that despite this mild singularity it is possible to write down a finite
 action integral of the non-commutative field thoery on $S^4$. 
NS-NS $B$ field background on $S^4$ which is associated with our Matrix $S^4$ 
configurations is also constructed. 
\end{abstract}
\newpage
\setlength{\baselineskip}{18pt}

\newcommand {\beq}{\begin{equation}}
\newcommand {\eeq}{\end{equation}}
\newcommand {\beqa}{\begin{eqnarray}}
\newcommand {\eeqa} {\end{eqnarray}}
\newcommand{\bm}[1]{\mbox{\boldmath $#1$}}
\newcommand{\Sq}{D(X)}
\newcommand{\al}{2\pi \alpha'}

\section{Introduction}
\hspace{5mm}
The fuzzy 4-sphere ($S^4$) was considered in \cite{CLT} to describe spherical 
Longitudinal 5-branes (L5-branes) in the context of the Matrix Theory 
\cite{BFSS}\cite{IIB} and to construct an example of finite 4d field 
theory\cite{GKP}\cite{Oconner}. 
By using the $n$-fold symmetric tensor product representation 
of the 4d gamma matrices, $N=(n+1)(n+2)(n+3)/6$ dimensional Matrix L5-brane 
configuration was constructed. There is, however, a problem in describing the 
fluctuating L5-branes.  Later it was concluded that `fuzzy $S^4$' is 6 
dimensional, {\it i.e.}, a fuzzy $S^2$ fibre bundle over a non-associative 
$S^4$.\cite{HoRangoolam}\cite{Rangoolam}

In the above works the assumption of the $SO(5)$ invariance of the fuzzy 
$S^4$ algebra was crucial. 
What underlies the non-commutativity is, however, the symplectic form or 
the NS-NS $B$ field background and there are no rotationally invariant 
symplectic forms on $S^m$ 
except for $S^2$. Actually, to construct $SO(5)$ invariant $B$ field 
background it is necessary to introduce an extra internal space and $SU(2)$ 
gauge symmetry.\cite{CNYang}\cite{ZH} 

In this paper we will propose a new non-commutative $S^4$ algebra. 
We do {\it not} assume $SO(5)$ invariance of the algebra and do  {\it not}
introduce an extra internal space. Our $S^4$ algebra has $SO(3) \otimes SO(2)$ 
symmetry, the subgroup of $SO(5)$. It is not a Lie algebra.  This algebra has 
$N=2(2j+1)$ dimensional irreducible representations $(j=1/2,1,3/2,...)$
and can be derived from a Matrix theory action as equations of motion. 
These representations are given in the form of a tensor product of two 
representations of $SO(3)$. Because $SO(3)$ is the defining algebra of the 
non-commutative $S^2$ \cite{Madore} and the matrix $\leftrightarrow$ function 
correspondence for $S^2$ is well known\cite{Presnejder}\cite{HNT}, the 
non-commutative geometry of $S^4$ can be realized 
in terms of the non-commutative structures on two $S^2$'s. We can  define the 
functions and the non-commutative product $\star$ on $S^4$ in terms of those 
on $S^2$. The formula for the special case of $N=4$ dim representation  
will be worked out explicitly. 
Actually, the topologies of $S^4$ and $S^2\otimes S^2$ 
are different and there appears mild singularity (indeterminateness) on a 
circle on $S^4$. We will, however, show that this singularity is mild enough 
such that a finite action integral can be written down. 
This situation is reasonable from the matrix $\leftrightarrow$ function 
correspondence: in the Matrix theory there is no singularity in a 
finite-size matrix. We will also show that $S^4$ constructed in the 
$S^2 \otimes S^2$ parametrization has a twofold structure.

The structure of this paper is as follows. In sec.2 we will define a Matrix 
theory action for spherical 4-brane and derive the non-commutative 
$S^4$ algebra. We will show that the $SO(5)$ symmetry of $S^4$ cannot be kept 
intact in the algebra. We find Matrix configurations with dimension 
$N=2(2j+1)$ $\ \ (j=1/2,1, 3/2,...)$. 
The corresponding non-commutative $S^4$ will be called $(S^4)_j$. 
In sec.3 a non-commutative product $\star$ for non-commutative field theories 
on $(S^4)_j$ will be constructed in terms of the non-commutative product 
$\ast$ on $S^2$. We point out that our $S^4$ has a twofold structure. 
Functions on $(S^4)_j$ will be also identified. In sec.4 the non-commutative 
product for $(S^4)_{1/2}$ will be presented explicitly and in sec.5 we
present the NS-NS $B$ field background corresponding to the Matrix 
configurations obtained in sec.2. In sec.6  discussions are presented.
In appendix A we give the multiplication rule of the functions on 
$(S^4)_{1/2}$.  In appendix B the coefficient functions $L^{(n)}$ which appear
in the expression of the non-commutative product $\star$ 
on $(S^4)_{1/2}$ are presented.

Throughout this paper we will let the initial lowercase romans 
$a$, $b$, $c,...$ to take values $1,2,3$ and the middle romans $i,j,..=
4,5$, while the capitals $A,B,..$ will run from 1 to 5.

\section{The Algebra for Non-commutative $S^4$}
\hspace{5mm}
In this section we will propose the algebra of the coordinates of the 
non-commutative $S^4$. At the beginning we will assume $SO(5)$ invariance 
of the algebra. We will, however, shortly find this is not the case. 
Because the invariant tensors are $\delta_{AB}$ and $\epsilon_{ABCDE} $, 
the algebra of the coordinates will be given, up to a multiplicative 
constant, by
\begin{equation}
[\hat{X}^A,\hat{X}^B]=\epsilon_{ABCDE} \ \hat{X}^C \ \hat{X}^D \ 
\hat{X}^E. \qquad (A,B,..=1,2,..,5)
\label{S4algebra}
\end{equation}
Here  $\hat{X}^A$'s are finite-dimensional hermitian matrices and 
$\epsilon_{ABCDE}$ is the Levi-Civita symbol. One can show that the condition
\begin{equation}
(\hat{X}^A)^2-C^2 \ \bm{1}=0
\label{sphere}
\end{equation}
commutes with $\hat{X}^A$'s, where $C$ is a scalar and $\bm{1}$ an identity 
matrix.
 
The algebra (\ref{S4algebra}) can be derived as equations of motion from the 
following action.
\begin{equation}
\int dt \ Tr \left\{\frac{1}{2R_M} \ (D_0 \ \hat{X}^A)^2 + \frac{1}{4}\left(
[\hat{X}^A,\hat{X}^B]-\epsilon_{ABCDE} \ \hat{X}^C \ \hat{X}^D \ \hat{X}^E 
\right)^2 \right\}
\label{M_action}
\end{equation} 
Here $D_0$ is the covariant derivative $\partial_t+i \ [\hat{A}_0, \ ]$ and 
$R_M$ the radius of the M circle. This may be regarded as the bosonic 
part of the M-atrix theory in some background.\footnote{The bosonic part of 
the action for Matrix theory in the pp-wave background has this 
form.\cite{BMN}}  
Here we omit other bosonic coordinates, $\hat{X}^6,...,\hat{X}^9$.
The potential is positive semi-definite and takes the minimal value when 
$\hat{X}^A$ satisfies the algebra (\ref{S4algebra}).
The time-independent solution to (\ref{S4algebra}) solves 
the equation of motion for the action (\ref{M_action}).

A representation of (\ref{S4algebra}) in terms of 4 $\times$ 4 matrices 
is given by tensor products of Pauli matrices.\footnote{We attached a 
subscript $0$ to $\hat{X}^A$'s to show that these are particular matrices 
that satisfy algebra (\ref{S4algebra}). }
\begin{eqnarray}
\hat{X}_0^1&=& \frac{1}{3} \sigma_3 \otimes \sigma_1, \qquad 
\hat{X}_0^2= \frac{1}{3} \sigma_3 \otimes \sigma_2, \nonumber \\
\hat{X}_0^3&=& \frac{1}{3} \sigma_3 \otimes \sigma_3, \qquad
\hat{X}_0^4= \frac{1}{3} \sigma_1 \otimes \bm{1}_2 , \nonumber \\
\hat{X}_0^5&=& \frac{1}{3} \sigma_2 \otimes \bm{1}_2 
\label{4dimrep}
\end{eqnarray}
$\bm{1}_2$ stands for 2 $\times$ 2 identity matrix.
These are four dimensional gamma matrices. 
It is then natural to try $N=2(2j+1)$ dimensional representation by 
replacing one of the two Pauli matrices by a spin-$j$ representation 
of $SO(3)$, $\ T_{(j)}^a$:
$\ \left([T_{(j)}^a, T_{(j)}^b]= \ i \ \epsilon_{abc} \ T_{(j)}^c \right)$.
\begin{eqnarray}
\hat{X}_0^1&=& \frac{2}{3} \sigma_3 \otimes T^1_{(j)}, \qquad 
\hat{X}_0^2= \frac{2}{3} \sigma_3 \otimes T^2_{(j)}, \nonumber \\
\hat{X}_0^3&=& \frac{2}{3} \sigma_3 \otimes T^3_{(j)}, \qquad
\hat{X}_0^4= \frac{1}{3} \sigma_1 \otimes \bm{1}_{2j+1}, \nonumber \\
\hat{X}_0^5&=& \frac{1}{3} \sigma_2 \otimes\bm{1}_{2j+1}
\label{Ndimrep}
\end{eqnarray}
Indeed, these $\hat{X}_0^A$'s satisfy (\ref{S4algebra}) {\em except} for the 
following commutator.
\begin{equation}
 \ [\hat{X}^4, \ \hat{X}^5] \ = \ \frac{3}{4j(j+1)} \ \epsilon_{45CDE} \ 
\hat{X}^C \ \hat{X}^D \ \hat{X}^E
\end{equation}
Although this equation breaks the symmetry $SO(5)$ down to 
$SO(3) \otimes SO(2)$, this algebra is a natural modification of 
(\ref{S4algebra}).
To recover the $SO(5)$ symmetry we may try rescaling $\hat{X}_0^a$ $\ 
(a=1,2,3)$
by a constant $\alpha$ and $\hat{X}_0^i$ $\ (i=4,5)$ by  $\beta$, respectively:
\begin{eqnarray}
\hat{X}_0^1&=& \frac{2}{3} \alpha \ \sigma_3 \otimes T^1_{(j)}, \qquad 
\hat{X}_0^2= \frac{2}{3} \alpha \ \sigma_3 \otimes T^2_{(j)}, \nonumber \\
\hat{X}_0^3&=& \frac{2}{3} \alpha \ \sigma_3 \otimes T^3_{(j)}, \qquad
\hat{X}_0^4= \frac{1}{3} \beta \ \sigma_1 \otimes \bm{1}_{2j+1}, \nonumber \\
\hat{X}_0^5&=& \frac{1}{3} \beta \ \sigma_2 \otimes\bm{1}_{2j+1}
\label{Ndimrepr}
\end{eqnarray}

We then obtain the algebra
\begin{eqnarray}
\ [\hat{X}^a, \ \hat{X}^b] &=& \frac{\alpha}{\beta^2} \ \epsilon_{abcij}
\left(\hat{X}^c \ \hat{X}^i \
\hat{X}^j-\hat{X}^i \ \hat{X}^c \ \hat{X}^j+\hat{X}^i \ \hat{X}^j \ \hat{X}^c \right), \nonumber \\
\ [\hat{X}^a, \ \hat{X}^i] &=& \frac{1}{\alpha} \ \epsilon_{aibcj}
\left(\hat{X}^b \ \hat{X}^c \
\hat{X}^j-\hat{X}^b \ \hat{X}^j \ \hat{X}^c+\hat{X}^j \ \hat{X}^b \ \hat{X}^c \right), \nonumber \\
\ [\hat{X}^4, \ \hat{X}^5] &=& \frac{3\beta^2}{4j(j+1)\alpha^3} \ 
\epsilon_{45abc}\ \hat{X}^a \ \hat{X}^b \ \hat{X}^c. 
\label{S4algebra2}
\end{eqnarray}
(Throughout this paper $a, b, .. =1,2,3$ and $i,j,..=4,5$, 
while $A,B,..=1,2,..,5$. )
When we try to equate the prefactors on the RHS, we find that the equations 
$\alpha/\beta^2=1/\alpha=3\beta^2/(4j(j+1)\alpha^3)$ for $\alpha$, $\beta$ do 
not have a solution except for the case $j=1/2$. Hence rescaling of 
$\hat{X}_0^A$'s does not save the symmetry; we conclude that we must 
abandon $SO(5)$ symmetry.

Although $\alpha^{-2}\ (\hat{X}^a)^2+ \beta^{-2} \ (\hat{X}^i)^2 $ does 
{\em not} commute with $\hat{X}^A$ in the algebra (\ref{S4algebra2}), the 
configuration (\ref{Ndimrepr}) {\em does} satisfy the following equation.
\begin{equation}
\alpha^{-2}\ (\hat{X}_0^a)^2+ \beta^{-2} \ (\hat{X}_0^i)^2= \frac{2}{9}\ (2j+1)^2
 \ \bm{1}
\label{X2}
\end{equation}
Since (\ref{S4algebra2}) is not a Lie algebra, this is possible. 
These matrices do not commute with each other and are not decoupled; the
matrices (\ref{Ndimrepr}) may be called a Matrix $S^4$ configuration.\footnote{
For $\alpha \neq \beta$ it may be more pertinent to call this an ellipsoid.
See footnote 4.  }
We will denote this configuration as $(S^4)_{j}$. Its dimension $N=2(2j+1)$ is
extending to infinity.  The structures of $\hat{X}_0^{4,5}$ in 
(\ref{Ndimrepr}) 
are quite simple compared to those of $\hat{X}_0^{1,2,3}$. Especially, the 
eigenvalues of $\hat{X}_0^{4,5}$ are $\pm \beta$, while those of 
$\hat{X}_0^{1,2,3}$ are randomly distributed over the range between 
$-(2j/3) \alpha$ and $(2j/3) \alpha $. In sec.6 we will consider more 
general matrices. For these the eigenvalues of $\hat{X}_0^{1,2,3,4,5}$ are 
all randomly distributed. 

We expect that when the size of the matrices becomes large, the matrices 
$\hat{X}^A_0$ will commute and the classical geometry of $S^4$ will be 
recovered as in the case of fuzzy $S^2$. To show that this is really the case 
we must replace the Pauli marices in the tensor products in (\ref{Ndimrepr}) 
to matrices of an arbitrary representation (spin $j'$). 
In sec.6 we will discuss the commutative limit.

Finally, the Matrix theory action for spherical 4-branes is given by 
\begin{eqnarray}
S &=& \int dt \ Tr \left\{\frac{1}{2R_M} \ (D_0 \ \hat{X}^A)^2 \right. 
\nonumber \\
&&  + \frac{1}{4}\left(
[\hat{X}^a,\hat{X}^b]-\frac{\alpha}{\beta^2} \ \epsilon_{abCDE} \ 
\hat{X}^C \ \hat{X}^D \ \hat{X}^E \right)^2   \nonumber \\
&&  + \frac{1}{2}\left(
[\hat{X}^a,\hat{X}^i]-\frac{1}{\alpha} \ \epsilon_{aiCDE} \ \hat{X}^C \ 
\hat{X}^D \ \hat{X}^E \right)^2  \nonumber \\
&& \left. + \frac{1}{2}\left(
[\hat{X}^4,\hat{X}^5]-\frac{3\beta^2}{4j(j+1)\alpha^3} \ \epsilon_{45CDE} \ 
\hat{X}^C \ \hat{X}^D \ \hat{X}^E \right)^2 \right\}.  
\label{M_action2}
\end{eqnarray} 
By shifting $\hat{X}^A \rightarrow \hat{X}_0^A + \hat{A}^A$ in this action,
where $\hat{X}_0^A$ is the Matrix $S^4$ configuration (\ref{Ndimrepr}), 
we obtain a matrix form of the action of the non-commutative gauge theory 
on $S^4$. 

\section{Non-commutative Product on $(S^4)_j$}
\hspace{5mm}
In this section we will construct a non-commutative product on $S^4$ 
corresponding to the Matrix configuration (\ref{Ndimrepr}) in terms of the 
product on $S^2$. 

\subsection{Non-commutative $S^2$}
\hspace*{5mm}
Two dimensional fuzzy sphere is defined by the $SO(3)$ algebra\cite{Madore} 
\begin{equation}
\ [ \hat{X}^a, \ \hat{X}^b ] = i \epsilon_{abc} \ \hat{X}^c \qquad 
(a,b,..=1,2,3).
\label{so3}
\end{equation}
This algebra can be realized in the space of functions on the sphere, where
$\hat{X}^a$ is represented by a variable $x^a$ and the multiplication rule 
for the functions on the sphere is defined by a noncommutative product $\ast$.
\footnote{We will use a symbol $\ast$ for the non-commutative product 
on $S^2$ and $\star$ for that on $S^4$.}
\begin{eqnarray}
f(x) \ast g(x)
& =& f(x) \ g(x)  +\sum_{m=1}^{\infty} \lambda^m \ C_m(\lambda)
J_{a_1 b_1}(x) \cdots J_{a_m b_m}(x) \nonumber \\
&& \qquad  \qquad \quad \qquad \times \partial_{a_1} \cdots \partial_{a_m} 
\ f(x) \ \partial_{b_1} \cdots \partial_{b_m} \ g(x) 
\label{star}
\end{eqnarray}
Here $f(x)$ and $g(x)$ are functions of $x^a/r$. $(r=\sqrt{(x^a)^2})$  
$C_m(\lambda)$ and $J_{ab}(x)$ are defined by
\begin{equation}
C_m(\lambda)=\frac{\lambda^m}{m! (1-\lambda)(1-2\lambda) 
\cdots (1-(m-1)\lambda)}, 
\end{equation}
\begin{equation}
J_{ab}(x)=r^2 \delta_{ab}-x^a \ x^b +i \ r \epsilon_{abc} \ x^c.
\end{equation}
By using this product we obtain the star-commutator. 
\begin{equation}
 \ [ x^a, \ x^b]_{\ast} \equiv x^a \ast x^b-x^b \ast x^a = \ 2  i \ 
\lambda \  r \ \epsilon_{abc} \ x^c
\label{so3x}
\end{equation}
By comparing (\ref{so3}) and (\ref{so3x}) we find the correspondence
$\hat{X}^a \leftrightarrow \frac{x^a}{2 \lambda r}$.  
For spin-$j$ representation of $SO(3)$ the quadratic Casimir operator takes 
a value $j(j+1)$ $\ (j=1/2,1,3/2,...)$. By using 
\begin{equation}
x^a \ast x^a = (1+2\lambda) r^2
\end{equation}
we obtain two values $\lambda=1/(2j)$ and $\lambda=-1/(2j+2)$. 

The first value $\lambda=1/(2j)$ corresponds to a finite set of functions on 
the sphere. The set of the polynomials of $x^a/r$, spherical harmonics, up to 
order $2j$ are closed under the multiplication defined by (\ref{star})
and the number of spherical harmonics is given by 
$\sum_{\ell=0}^{2j} \ (2\ell+1)=(2j+1)^2$.  This is equal to the number of 
independent components of ($2j+1$) $\times$ ($2j+1$) hermitian matrix. 
In this case the summation over $m$ in (\ref{star}) must be terminated at 
$m=2j$.

For the second value $\lambda=-1/(2j+2)$ the spherical harmonics to 
arbitrary orders are produced under multiplication of $x^a/r$'s and the number 
of independent functions is infinite. 

In this paper we will restrict our attention to a finite set of functions on 
the non-commutative $S^4$ and henceforth concentrate on the first value 
$\lambda=1/(2j)$.  
In this case we find the matrix $\leftrightarrow$ function correspondence, the 
correspondence between  the generator in the spin $j$ representation,
$\hat{X}^a=T_{(j)}^a$, and $x^a$.
\begin{equation}
T_{(j)}^a \ \leftrightarrow \ j \ \frac{x^a}{r}
\label{correspondence}
\end{equation}
Both sides satisfy the same $SO(3)$ algebra. 

\subsection{$S^2 \otimes S^2$ parametrization of $S^4$}
\hspace*{5mm} 
We will now apply the correspondence (\ref{correspondence}) 
to the $S^4$ configuration (\ref{Ndimrepr}). 
These matrices are written as tensor products of two matrices, whose 
dimensions are $2$ and $2j+1$, respectively. $2 \times 2$ matrices are spanned
by $\bm{1}$ and $\sigma_a$, while $(2j+1) \times (2j+1)$ matrices are spanned 
by $\bm{1}$ and the products of $T^a_{(j)}$'s. 
By matrix $\leftrightarrow$ function correspondence these matrices are 
realized by polynomials on $S^2$. Then we have to introduce two $S^2$'s.  
The coordinates of each $S^2$ are denoted by 
$(x^1,x^2,x^3)$ and $(y^1,y^2,y^3)$, respectively. The radii of the two $S^2$'s
are $r=\sqrt{(x^a)^2}$ and  $\rho=\sqrt{(y^a)^2}$. 
Then in (\ref{Ndimrepr}) we replace $\frac{1}{2}\sigma_a$ by 
$x^a/2r$ and $T^a_{(j)}$ 
by $j \ y^a/\rho$. 
\begin{eqnarray}
\hat{X}_0^1 \leftrightarrow X^1 &=& R \ \frac{x^3}{r} \frac{y^1}{\rho}, \qquad 
\hat{X}_0^2 \leftrightarrow X^2 = \ R \ \frac{x^3}{r} \frac{y^2}{\rho}, 
\nonumber \\
\hat{X}_0^3 \leftrightarrow X^3 &=& R \ \frac{x^3}{r} \frac{y^3}{\rho}, \qquad
\hat{X}_0^4 \leftrightarrow X^4 = \ R \ \frac{x^1}{r}, \nonumber \\
\hat{X}_0^5 \leftrightarrow X^5 &=& R \ \frac{x^2}{r}
\label{S4coordinates}
\end{eqnarray}
Here we set $\alpha=3R/2j$ and $\beta=3R$. These relations give a 
transformation from $\left(\frac{x^a}{r},\frac{y^b}{\rho}\right)$ 
to $\left(\frac{X^A}{R}\right)$.
Because we can check that\footnote{In this case (\ref{X2}) will represent 
an ellipsoid for $j \neq 1/2$. If we made a different choice for 
$\alpha$ and $\beta$, the manifold described by the coordinate $X^A$ would  
in turn be deformed into an ellipsoid.   
For the discussion of this paper this choice of $\alpha$ and $\beta$ 
is just for convenience.}
\begin{equation}
\sum_{A=1}^5 \ (X^A)^2 =R^2, \qquad \mbox{(ordinary product)}
\end{equation}
 (\ref{S4coordinates}) yields a map from $S^2 \otimes S^2$
to $S^4$. 
In fact this map gives a double cover of $S^4$. Actually, two points on 
$S^2 \otimes S^2$, 
\[ P=((x^1,x^2,x^3)/r,(y^1,y^2,y^3)/\rho) \ \  \mbox{and} \ \  
P'=((x^1,x^2,-x^3)/r,(-y^1,-y^2,-y^3)/\rho)\]
map onto a same point on $S^4$. 
Therefore we can divide the first $S^2$ into the upper and lower hemispheres, 
$S^2_+$ ($ \ x^3 \geq 0$) and $S^2_-$ ($ \ x^3 \leq 0$), each corresponding 
to a single $S^4$. The boundary points $(x^3=0)$, which constitute 
$S^1 \otimes S^2$, will be mapped onto a great circle ($S^1$),  
\begin{equation}
{\cal C}=\{(X^1,X^2,X^3,X^4,X^5)| \ X^1=X^2=X^3=0, \ (X^4)^2+(X^5)^2=R^2 \}.
\label{circle}
\end{equation}
Consequently, the inverse map $S^4 \ \rightarrow \ S^2_+ 
\otimes S^2$, 
\begin{eqnarray}
x^1(X) &=& r \ X^4 /R, \qquad x^2(X) = r \ X^5/R, \qquad 
 x^3(X) =  r \ D(X)/R, \nonumber  \\
y^a(X) &=&  \rho \ X^a/D(X),
\label{inversemap}
\end{eqnarray}
is multi-valued (indeterminate) on the circle. The image of each point on 
${\cal C}$ is $S^2$.  Here $D(X)$ is defined by
\begin{equation}
D(X) \equiv \sqrt{(X^1)^2+(X^2)^2+(X^3)^2}.
\label{DX}
\end{equation}
The other inverse map $S^4 \ \rightarrow \ S^2_- \otimes S^2$, which is 
obtained from (\ref{inversemap}) by the replacement $D(X)\rightarrow -D(X)$, 
is also indeterminate on ${\cal C}$.

The above analysis shows that in the $S^2 \otimes S^2$ parametrization 
the manifold $S^4$ which corresponds to the Matrix 
configuration (\ref{Ndimrepr}) is  doubled. The two $S^4$'s are 
attached on ${\cal C}$. The non-commutative structures on each $S^4$ are 
related by $D(X)\leftrightarrow -D(X)$. This fact suggests us to assign 
distinct signs to $D(X)$ on each sheet of the doubled $S^4$:
\begin{equation}
D(X) \equiv \left\{ \begin{array}{cc}
                + \sqrt{(X^a)^2} & \mbox{ on the first $S^4$} \\
                 - \sqrt{(X^a)^2} & \mbox{ on the second $S^4$}
                               \end{array} \right.
\end{equation}

\subsection{Non-commutative Product}
\hspace*{5mm}
Let us now construct the noncommutative product on $(S^4)_j$. 
Because a function on $S^2 \otimes S^2$ does not depend on $r$ and $\rho$, 
this function does not depend on $R$, either, and this can also be  
regarded as that  on $S^4$. To write down the non-commutative product on 
$S^4$ we must regard $R$ as a function of $x^a$ and $y^a$. To keep $SO(3)$ 
symmetry of $S^2$'s intact we will take $R=R(r,\rho)$. Then 
(\ref{inversemap}) can also be regarded as a transformation of six variables 
$(x^a,y^b) \ \rightarrow \ (X^A,R)$. By combining the non-commutative 
products (\ref{star}) for the two $S^2$'s we define the noncommutative 
product of functions, $F(X)$ and $G(X)$, on $(S^4)_j$.
\begin{eqnarray}
&&F(X) \star  G(X) \nonumber \\
&&= F(X)  G(X) + (r^2\delta_{ab}-x^ax^b+i \ r \  \epsilon_{abc}x^c)
\frac{\partial F(X)}{\partial x^a} \ 
\frac{\partial G(X)}{\partial x^b} \nonumber \\
&& \qquad +\lambda (\rho^2\delta_{ab}-y^ay^b+i \ \rho \ \epsilon_{abc}y^c)
\frac{\partial F(X)}{\partial y^a} \ \frac{\partial G(X)}{\partial y^b} 
\nonumber \\ && \qquad +\lambda (r^2\delta_{ab}-x^ax^b+i \ r \ 
\epsilon_{abc}x^c)
(\rho^2\delta_{de}-y^dy^e+i \ \rho \  \epsilon_{def}y^f)
\frac{\partial^2 F(X)}{\partial x^a \partial y^d} \ 
\frac{\partial^2 G(X)}{\partial x^b \partial y^e} \nonumber \\
&& \qquad + \cdots
\label{product1}
\end{eqnarray}
Here $\lambda=1/2j$ and terms with higher orders of $\lambda$ are not written 
explicitly.  This is still expressed in terms of $x^a$, $y^a$, not by 
$X^A$, and implicit. Later we will write down the product for the 
$j=1/2$ case more explicitly.

\subsection{Functions on Non-commutative  $(S^4)_j$  }
\hspace*{5mm}
The functions on $(S^4)_j$ are given by the products of the functions on the two 
$S^2$'s.  Functions on the first $S^2$ are spanned by $1$ and 
$\{x^a/r, \ (a=1,2,3)\}$.  Those on the second $S^2$ are spanned by $1$ and 
$\{y^{a_1} \cdots y^{a_n}/\rho^n$, $(1 \leq n \leq 2j, 
\ a_1, .., a_n=1,2,3)\}$. 
Then those on $(S^4)_j$ are given by linear combinations of 
\begin{equation}
 1, \qquad x^a/r, \qquad y^{a_1} \cdots y^{a_n}/\rho^n, \qquad x^a y^{a_1} 
\cdots y^{a_n}/r\rho^n. 
\label{funcsonS}
\end{equation}
These functions can be expressed in terms of $X^A$'s by the transformation 
(\ref{inversemap}).  
\begin{eqnarray}
&& 1, \qquad  X^i/R, \qquad D(X)/R, \qquad X^{a_1} \cdots X^{a_n}/D(X)^n, 
\nonumber \\ &&
X^i X^{a_1} \cdots X^{a_n}/R D(X)^n, \qquad 
X^i X^{a_1} \cdots X^{a_n}/R D(X)^n, \nonumber \\ && X^{a_1} 
\cdots X^{a_n}/R D(X)^{n-1}
\label{funcsonS4}
\end{eqnarray}
The number of independent functions is $4 \cdot (2j+1)^2=\{2  (2j+1)\}^2$, 
which is a square
of an integer. This is the reason why the functions on $(S^4)_j$ correspond to 
$2 (2j+1) \times 2 (2j+1) $ hermitian matrices. All the functions are, 
however, not polynomials. 
 
The spherical harmonics $Y_{\ell_1 \ell_2 \ell_3 m}$ on $S^4$ are labeled by 
integers $\ell_1 (=0,1,..)$, $\ell_2 (=0,1,..,\ell_1)$, 
$\ell_3 (=0,1,..,\ell_2)$, $m (=0,\pm1,.., \pm \ell_3)$. 
These are $\ell_1$-th order polynomials and the total number of polynomials up
to order $\ell$ is given by 
\begin{equation}
\sum_{\ell_1=0}^{\ell} \ \sum_{\ell_2=0}^{\ell_1} \ \sum_{\ell_3=0}^{\ell_2} \ 
(2\ell_3+1) = \frac{1}{12} \ (\ell+1) \ (\ell +2)^2 \ (\ell+3). 
\label{el}
\end{equation}
This is not a square of an integer. 
With any choice of $\ell$ (\ref{el}) cannot be made equal to $\{2(2j+1)\}^2$, 
the number of independent functions on $S^4$. 
Indeed this fact makes it difficult 
to establish an isomorphism between matrices and polynomials on $S^4$ equipped
with an associative multiplication rule. \cite{HoRangoolam}\cite{Rangoolam}

Our construction avoids this problem by introducing non-polynomial functions 
in (\ref{funcsonS4}). These functions  are, however, not 
well-defined at all points on $S^4$. 
The map $S^4 \rightarrow S^2_+ \otimes S^2$ (\ref{inversemap}) is
multi-valued on the circle (\ref{circle}) and the functions (\ref{funcsonS4}) 
are also indeterminate.  Actually these functions are defined on 
$S^2 \otimes S^2$ and such a singularity is anticipated. In ordinary 
commutative field theory such a singularity of the fields is not allowed. 
Derivatives of the fields will become more singular. On the other hand in 
non-commutative field theory this singulatity is, however, not serious. 
First of all this discontinuity is finite and the functions (\ref{funcsonS4}) 
are integrable on $S^4$. In addition, in non-commutative field theories the 
derivative of a function $F(X)$ is given by the commutator 
$\nabla_A \ F(X) = \frac{i}{R} [ X^A, F(X)]_{\star }=\frac{i}{R} \ (X^A \star 
F(X) - F(X) \star X^A) $, which can be expressed as a linear combination of 
the functions (\ref{funcsonS4}).
Because the non-commutative product contains the derivatives of functions, 
each term in the product may be singular on the circle. Let us, however, 
define the non-commutative product of the functions on the circle by a limit
from outside the circle. In this case, when the product $\star$ is properly 
constructed, it is arranged such that those singularities cancel out 
completely to leave only the indeterminateness on ${\cal C}$. 
Therefore in spite of the indeterminateness of the functions on the circle
(\ref{circle}) the product and derivatives of functions are \lq well-defined'
and we can write down a finite action integral. 
This is sufficient for our purpose. 

This point is also evident from 
the matrix $\leftrightarrow$ function correspondence:  
the Matrix configuration (\ref{Ndimrepr}) does not 
have any singurality. The trace of a finite-size matrix is finite. 

We conclude that we can construct non-commutative field 
theories on $S^4$ at the expense of the indeterminateness of functions on 
the circle (\ref{circle}).
It still, however, remains a possibility that a different 
method other than our $S^2 \otimes S^2$ parametrization might lead to 
non-singular, non-commutative product and functions. 

To complete the matrix-function correspondence we need to define the 
integration measure over $S^4$.  This is again induced by the integration
measures over the two $S^2$'s, which are given by $\int dx^1 \ dx^2 /r x^3$ 
and $\int dy^1 \ dy^2 / \rho y^3$, respectively.  By using (\ref{inversemap}) 
we define the integration measure over $S^4$ by 
\begin{equation}
\int_{S^4} d^4 X \equiv \int \frac{dx^1 \ dx^2}{r \ x^3} \ \frac{dy^1 \ dy^2}
{\rho \ y^3}
=\int \frac{dX^1 \ dX^2 \ dX^4 \ dX^5}{R \ X^3 \ D(X)^2}.
\end{equation}
This does not coincide with the ordinary $SO(5)$ invariant measure.

\section{Non-Commutative Product On $(S^4)_{j=1/2}$} 
\hspace*{5mm}
We will work out the explicit form of the non-commutative product for the 
$j=1/2$ case. 
Although it is in principle possible to change variables 
from $(x^a,y^a)$ to $X^A$ in (\ref{product1}), we found it easier to construct the product 
which reproduces the multiplication rule of the functions (\ref{funcsonS4}).
In the $j=1/2$ case we have the following functions. 
\begin{eqnarray}
&&1, \qquad  X^i/R=x^{i-3}/r, \qquad D(X)/R=x^3/r, \qquad X^{a}/D(X)=y^a/{\rho},  \nonumber \\
&& X^{a}/R=x^3y^a/r\rho, \qquad X^i X^a/RD(X)=x^{i-3}y^a/r\rho
\label{funcs1onS4}
\end{eqnarray}
By using the multiplication rule of functions on $S^2$,
\begin{eqnarray}
&& 1 \ast  1=1, \qquad 1 \ast  \frac{x^a}{r}=\frac{x^a}{r}*1= \frac{x^a}{r}, 
\nonumber
 \\
&& \frac{x^a}{r} \ast  \frac{x^b}{r} = \delta_{ab}
+i \epsilon_{abc}\frac{x^c}{r}
\end{eqnarray}
and the similar equations for $y^a/\rho$, we obtain the
multiplication rule of the functions (\ref{funcs1onS4}). This is  presented 
in Appendix A.

Because the non-commutative product for $S^2$ in the $j=1/2$ representation 
contains at most first order derivatives of functions, we find that the product
for $(S^4)_j$ contains at most second order derivatives of functions.  
So we can assume the following form for the product. 
\begin{eqnarray}
F(X) \star G(X)
= && FG  \nonumber \\
& + & 
L^{(1)}_{a,b}
\frac{\partial F}{\partial X^a}
\frac{\partial G}{\partial X^b}
+
L^{(2)}_{i,j}
\frac{\partial F}{\partial X^i}
\frac{\partial G}{\partial X^j}
\nonumber \\
& + &
L^{(3)}_{a,i}
\frac{\partial F}{\partial X^a}
\frac{\partial G}{\partial X^i}
+
L^{(4)}_{i,a}
\frac{\partial F}{\partial X^i}
\frac{\partial G}{\partial X^a}
\nonumber \\
& + &
L^{(5)}_{ab,c}
\frac{\partial^2 F}{\partial X^a \partial X^b}
\frac{\partial G}{\partial X^c}
+
L^{(6)}_{a,bc}
\frac{\partial F}{\partial X^a}
\frac{\partial^2 G}{\partial X^b \partial X^c}
\nonumber \\
& + &
L^{(7)}_{ab,i}
\frac{\partial^2 F}{\partial X^a \partial X^b}
\frac{\partial G}{\partial X^i} 
+
L^{(8)}_{a,bi}
\frac{\partial F}{\partial X^a}
\frac{\partial^2 G}{\partial X^b \partial X^i}
\nonumber \\
& + &
L^{(9)}_{ai,b}
\frac{\partial^2 F}{\partial X^a \partial X^i}
\frac{\partial G}{\partial X^b}
+
L^{(10)}_{i,ab}
\frac{\partial F}{\partial X^i}
\frac{\partial^2 G}{\partial X^a \partial X^b}
\nonumber \\
& + &
L^{(11)}_{ai,j}
\frac{\partial^2 F}{\partial X^a \partial X^i}
\frac{\partial G}{\partial X^j} 
+
L^{(12)}_{a,ij}
\frac{\partial F}{\partial X^a }
\frac{\partial^2 G}{\partial X^i \partial X^j}
\nonumber \\
&+&
L^{(13)}_{i,aj}
\frac{\partial F}{\partial X^i}
\frac{\partial G}{\partial X^a \partial X^j}
+
L^{(14)}_{ij,a}
\frac{\partial^2 F}{\partial X^i \partial X^j}
\frac{\partial G}{\partial X^a}
\nonumber \\
&+&
L^{(15)}_{ij,k}
\frac{\partial^2 F}{\partial X^i \partial X^j}
\frac{\partial G}{\partial X^k} 
+
L^{(16)}_{i,jk}
\frac{\partial F}{\partial X^i}
\frac{\partial^2 G}{\partial X^j \partial X^k}
\nonumber \\
&+&
L^{(17)}_{ab,cd}
\frac{\partial^2 F}{\partial X^a \partial X^b}
\frac{\partial^2 G}{\partial X^c \partial X^d}
+
L^{(18)}_{ij,kl}
\frac{\partial^2 F}{\partial X^i \partial X^j}
\frac{\partial^2 G}{\partial X^k \partial X^l}
\nonumber \\
& + &
L^{(19)}_{ab,ci}
\frac{\partial^2 F}{\partial X^a \partial X^b}
\frac{\partial^2 G}{\partial X^c \partial X^i}
+
L^{(20)}_{ai,bc}
\frac{\partial^2 F}{\partial X^a \partial X^i}
\frac{\partial^2 G}{\partial X^b \partial X^c}
\nonumber \\
& + &
L^{(21)}_{ab,ij}
\frac{\partial^2 F}{\partial X^a \partial X^b}
\frac{\partial^2 G}{\partial X^i \partial X^j}
+
L^{(22)}_{ij,ab}
\frac{\partial^2 F}{\partial X^i \partial X^j}
\frac{\partial^2 G}{\partial X^a \partial X^b}
\nonumber  \\
&+&
L^{(23)}_{ai,bj}
\frac{\partial^2 F}{\partial X^a \partial X^i}
\frac{\partial^2 G}{\partial X^b \partial X^j}
\nonumber \\
& + &
L^{(24)}_{ai,jk}
\frac{\partial^2 F}{\partial X^a \partial X^i}
\frac{\partial^2 G}{\partial X^j \partial X^k}
+
L^{(25)}_{ij,ak}
\frac{\partial^2 F}{\partial X^i \partial X^j}
\frac{\partial^2 G}{\partial X^a \partial X^k}
\label{starjhalf}
\end{eqnarray}
Here $F(X)$ and $G(X)$ are arbitrary linear combinations of 
(\ref{funcs1onS4}). $L^{(n)}$  are suitable functions of $X^A$.  
These functions $L^{(n)}$ must be determined in such a way that 
(\ref{mt23}) is reproduced.  
In addition we must also require that $\sqrt{(X^A)^2}=R$ is a constant.
\begin{equation}
f\left(\sqrt{(X^A)^2}\right) \star  G(X) = G(X) \star  
f\left(\sqrt{(X^A)^2}\right) = f\left(\sqrt{(X^A)^2}\right) \ G(X), 
\end{equation}
where $f(R)$ is an arbitrary function of $R$, and $G(X)$ a linear combination 
of (\ref{funcs1onS4}). Because the non-commutative product of functions 
contains up to second order derivatives, it is sufficient to take 
$f(R)=R, \ R^2$.

There is ambiguity in the structure of $L^{(n)}$.
The origin of this ambiguity lies in the arbitrariness of the  
$r$, $\rho$ dependence of $R(r,\rho)$. 
One of the simplest solutions is presented in Appendix B.

In the non-commutative space the derivatives  of the fields are defined in 
terms of the star commutator with $X^A$.
\begin{equation}
 \nabla_A G(X)\equiv \frac{i}{R} \ [X^A,G(X)]_{\star}
\end{equation}
Explicitly for $(S^4)_{1/2}$, by using (\ref{starjhalf}), 
(\ref{L1}) we obtain
\begin{eqnarray}
\nabla_a G(X)&\equiv&
-2\frac{R}{\Sq}\epsilon_{abc}X^c\frac{\partial G}{\partial X^b}
-2\frac{1}{\Sq}\epsilon_{ij}X^j X^a\frac{\partial G}{\partial X^i} \nonumber \\
&&-\frac{R^2-\Sq^2}{R\Sq}(\epsilon_{acd}X^b+\epsilon_{bcd}X^a)X^d
\frac{\partial^2 G}{\partial X^b \partial X^c} \nonumber \\ &&
+\frac{2}{R}\Sq\epsilon_{abc}X^c X^i
\frac{\partial^2 G}{\partial X^b \partial X^i} \nonumber \\
&&-2i\epsilon_{ij}\epsilon_{abc}X^j X^c
\frac{\partial^2 G}{\partial X^b \partial X^i} \nonumber \\&&
+\frac{1}{\Sq}(\epsilon_{ik}X^j+\epsilon_{jk}X^i)X^k X^a
\frac{\partial^2 G}{\partial X^i \partial X^j},
\end{eqnarray}

\begin{eqnarray}
\nabla_i G(X) &\equiv& 
-2\Sq \epsilon_{ij}
\frac{\partial G}{\partial X^j}
-2\frac{1}{\Sq}\epsilon_{ij}X^j X^a
\frac{\partial G}{\partial X^a} \nonumber \\
&&-2\frac{1}{\Sq}\epsilon_{ij}X^j X^a X^b
\frac{\partial^2 G}{\partial X^a \partial X^b}
-2\Sq X^a \epsilon_{ij}
\frac{\partial^2 G}{\partial X^a \partial X^j} \nonumber \\
&&-i\frac{\Sq^2}{2R}(\epsilon_{ij}\epsilon_{kl}+\epsilon_{ik}\epsilon_{jl})X^l
\frac{\partial^2 G}{\partial X^j \partial X^k}
\end{eqnarray}

The field strength $F_{AB}(X)$ of the non-commutative gauge field $A_A(X)$ is 
defined by replacing the matrix $\hat{X}^A$ in 
\begin{equation}
\frac{i}{R^2} \ \left([\hat{X}^A, \hat{X}^B]-\frac{1}{3 R} \ 
\epsilon_{ABCDE} \ \hat{X}^C \ \hat{X}^D \ \hat{X}^E \right), 
\end{equation}
which appears in the action (\ref{M_action2}), 
by $X^A +R \ A_A$, and the matrix multiplication by $\star$. Here we write 
down only the expression for the $j=1/2$ case. We obtain, by setting 
$ \alpha= \beta= 3R$, 
\begin{eqnarray}
F_{AB}(X) &=& \nabla_A \ A_B - \nabla_B \ A_A +i \ 
[A_A, A_B]_{\star } 
-\frac{i}{3} \ 
\epsilon_{ABCDE} \ A_C \star  A_D \star A_E \nonumber \\
&&-\frac{i}{3R^2} \epsilon_{ABCDE} \ (X^C \star  X^D \star  A_E 
+ X^C \star  A_D \star  X^E + A_C \star  X^D \star  X^E) \nonumber \\
&&-\frac{i}{3R} \epsilon_{ABCDE} \ (X^C \star  A^D \star  A_E 
+ A^C \star  X_D \star  A^E + A_C \star  A^D \star  X^E). \nonumber \\ &&
\end{eqnarray}
The action integral is given by the usual form.
\begin{equation}
S_{\mbox{gauge theory}}=\int dt \ \int_{S^4} d^4 X \ \left( \frac{R^2}{2R_M} \  D_0 \ A_A \star  D_0 \ A_A
-\frac{R^4}{4} F_{AB} \star  F_{AB}\right)
\end{equation}

\section{$B$ Field Background}
\hspace*{5mm}
Although non-commutative geometry is not always related to the NS-NS $B$ field 
background\cite{Meyers}, the relation between the $B$ field and the 
non-commutativity is well established.\cite{SW} 
When the $B$ field two-form is closed, $dB=0$, then the inverse matrix 
$\alpha^{AB}$ $ \ (B_{AB} \ \alpha^{BC}={\delta_A}^C)$ 
defines a Poisson bracket, 
$\{F, G\}_{PB} = \frac{1}{2}\alpha^{AB} \ \partial_A F \ \partial_B G$, 
and one can 
construct an associative, non-commutative product $F \star G$ by means of 
perturbation in $\alpha$ and its derivatives.\cite{Kontsevich} 
In what follows we will construct 
a $B$ field background (or symplectic form) on $S^4$.

Let us introduce an $S^2 \otimes S^2$ parametrization    
(\ref{S4coordinates}) of $S^4$.
\begin{eqnarray}
X^1 &=& R \ \cos \theta_1 \ \sin \theta_2  \ \cos \varphi_2, 
\qquad X^2=R \ \cos \theta_1 \ \sin \theta_2 \ \sin \varphi_2, 
\nonumber \\
X^3 &=& R \ \cos \theta_1 \ \cos \theta_2, \qquad 
\qquad X^4 =R \ \sin \theta_1 \ \cos \varphi_1, 
\nonumber \\
X^5 &=& R \ \sin \theta_1 \ \sin \varphi_1
\label{S2S2}
\end{eqnarray}
Here $(\theta_1,\varphi_1)$ and $(\theta_2,\varphi_2)$ are polar 
coordinates of two unit spheres $(0 \leq \theta_i \leq \pi, \ 0 \leq \varphi_i
\leq 2\pi)$ corresponding to $x^a$ and $y^a$, respectively. We can check 
that $X^A$'s satisfy $(X^A)^2=R^2$. 

This parametrization gives a double cover of $S^4$.  The points $P=(\theta_1,
\varphi_1, \theta_2,\varphi_2)$ and $P'=(\pi-\theta_1,\varphi_1, \pi-\theta_2,
\varphi_2+\pi)$ are mapped onto a same point on $S^4$. Hence we must divide  
the first $S^2$ into an upper hemisphere, $S^2_+$ $(0 \leq \theta_1 \leq 
\pi/2)$ and a lower one, $S^2_-$ $(\pi/2 \leq \theta_1 \leq \pi)$.
We must also be careful with the boundary of the hemispheres  
$(\theta_1=\pi/2, 
\ 0 \leq \varphi_1 \leq 2\pi)$. 
For any values of $\theta_2$, $\varphi_2$ this is mapped onto a circle 
(\ref{circle}).

The $B$ field background which has the maximal symmetry of the two spheres is 
given by
\begin{equation}
B \equiv \frac{n_1}{2} \ \sin \theta_1 \ d \theta_1 \wedge d \varphi_1 
+ \frac{n_2}{2} \ \sin \theta_2 \  d \theta_2 \wedge d \varphi_2. 
\label{B}
\end{equation}
Here the fluxes are quantized as usual and $n_1$, $n_2$ are integers.  
This two-form is closed. Now let us remember that for $\theta_1 =\pi/2$ 
the coordinates $\theta_2$ and $\varphi_2$ become redundant in (\ref{S2S2}). 
Therefore actually, the $B$ field (\ref{B}) is not defined at all points 
on $S^4$.  In terms of $X^A$ we obtain 
\begin{eqnarray}
B &=& \frac{n_2}{4D(X)^3} \epsilon_{abc}X^a \ dX^b \wedge dX^c 
+\frac{n_1 }{4 \ R \ D(X)} \ \epsilon_{ij} \ dX^i \wedge dX^j \nonumber \\
&\equiv& \frac{1}{2} \ B_{AB} \ dX^A \wedge dX^B. 
\label{BX}
\end{eqnarray}
This has $SO(3) \otimes SO(2)$ symmetry. 
The components subject to the condition of tangential projection 
$B_{AB} \ X^A=0$ are  
\begin{eqnarray}
&&B_{ab}= \frac{n_2}{2D(X)^3} \ \epsilon_{abc} \ X^c,  \qquad B_{ij}= 
\frac{n_1}{2RD(X)} \ \epsilon_{ij}, \nonumber \\
&&B_{ai}=-B_{ia} =  \frac{n_1 }{2RD(X)^3} \ \epsilon_{ij} \ X^j \ X^a. 
\end{eqnarray}
These components are singular on ${\cal C}$.

The inverse matrix $\alpha^{AB}$ which satisfies $\alpha^{AB}\ B_{BC}=
\delta_{AC}-X^A X^C/R^2$ is then given by
\begin{eqnarray}
&&\alpha^{ab} = - \frac{2D(X)}{n_2} \ \epsilon_{abc} \ X^c,  \qquad 
\alpha^{ij}= - \frac{2D(X)^3}{n_1 \ R} \ \epsilon_{ij}, \nonumber \\
&&\alpha^{ai}=-\alpha^{ia} = - \frac{2D(X)}{n_1 R}\ \epsilon_{ij} \ X^j \  X^a.
\end{eqnarray} 
Now $\alpha^{AB}$ defines the Poisson bracket.
\begin{equation}
 \{F(X), G(X)\}_{PB} \equiv \frac{1}{2} \ 
\alpha^{AB} \ \partial_A \ F(X) \ \partial_B \ G(X)
\end{equation}
We can easily show that $\alpha^{AB}$ satisfies the condition of associativity.
\begin{equation}
\alpha^{AB} \ \partial_B \ (\alpha^{CD}/R^2)+ (\mbox{cyclic permutations in } A,C,D) \ =0
\end{equation}
This is also valid on the circle (\ref{circle}), because $\alpha^{AB}=\partial_A \ \alpha^{BC}=0$ on ${\cal C}$. 
In terms of $\alpha^{AB}$ we can define the non-commutative 
product.\cite{Kontsevich} 
\begin{eqnarray}
&& F(X) \star' G(X) \nonumber \\
& =& F(X)  G(X) + i \ \alpha^{AB} \ 
\partial_A F(X) \ \partial_B G(X) \nonumber \\&&
- \frac{1}{2} \alpha^{AB} \ \alpha^{CD}
\ \partial_A \partial_C F(X) \ \partial_B \partial_D G(X)  \nonumber \\
&& - \frac{1}{3} \ \alpha^{AB} \ \left(\partial_B \ \alpha^{CD} 
\right) \ \left\{ \partial_A \partial_C F(X) \ \partial_D G(X) - 
\partial_C F(X) \ \partial_A \partial_D G(X) \ \right\} \nonumber \\
&&  +\cdots
\label{starpm}
\end{eqnarray}
The drawback of  this approach is that it is difficult to obtain an explicit 
expression of the non-commutative product in a closed form. From higher 
order terms  symmetric 
terms like $S^{AB} \ \partial_A F \ \partial_B G \quad (S^{BA}=S^{AB})$ will 
appear and even the coefficient function of the anti-symmetric term, 
$\alpha^{AB}(X) \ \partial_A F(X) \ \partial_B G(X)$, will be modified. 
In view of the symmetry of $B$ \ we, however, expect this product to be 
related to the Matrix configuration (\ref{Ndimrepr}) and 
the product (\ref{star}) obtained in sec.3. To establish this connection more
investigation will be necessary.
Because there are two integers, $n_1$, $n_2$, in (\ref{BX}), we suspect  
there will be more representations other than (\ref{Ndimrepr}).  
Matrix configuration  
(\ref{Ndimrepr}) will correspond to $n_1=1$ and  $n_2=2j$.

\section{Discussions}
\hspace*{5mm}
In this paper we have explicitly constructed Matrix 4-brane configurations 
(\ref{Ndimrepr}) and then defined the non-commutative $S^4$ algebra 
(\ref{S4algebra2}) and Matrix theory action (\ref{M_action2}). It turned out 
that the algebra is not invariant under $SO(5)$ but only under 
$SO(3) \otimes SO(2)$ and that the algebra is not even a Lie algebra. 
These Matrix configurations take the forms of tensor products of $2\times 2$
matrix and $(2j+1) \times( 2j+1)$ one. ($j=1/2,1,3/2,\ldots$)  
Then it is natural to expect more representations which depend on a  
parameter other than $j$. 
Actually, by simply replacing the Pauli matrices $\sigma_a$ 
in (\ref{Ndimrepr}) by $2 \ T_{(j')}^a$ we obtain 
\begin{equation}
\hat{X}_0^a=\frac{\alpha}{j'(j'+1)} \ T_{(j')}^3 \otimes T_{(j)}^a, 
\qquad \hat{X}_0^i= \frac{\beta}{2j'(j'+1)} \ T_{(j')}^{i-3} \otimes 
\bm{1}_{2j+1},
\label{extend}
\end{equation}
($a=1,2,3, \quad i=4,5$).  
While we still have $SO(3) \otimes SO(2)$ symmetry, it turns out that 
the algebra must be modified and the RHS of the algebra is not polynomials. 
\begin{eqnarray} 
\ [ \hat{X}^a, \hat{X}^b] &=& \frac{2\alpha}{\beta^2} \ j'(j'+1) \ 
\epsilon_{abc} \ \left( \hat{X}^c \ [\hat{X}^4,\hat{X}^5]+ [\hat{X}^4,
\hat{X}^5] \ \hat{X}^c \right), \nonumber \\
\ [ \hat{X}^a, \hat{X}^i] &=& \frac{j'(j'+1)}{\alpha} \ 
\epsilon_{abc} \ \epsilon_{ij} \ \left( \hat{X}^b \ [\hat{X}^c,\hat{X}^j]
- [\hat{X}^b,\hat{X}^j] \ \hat{X}^c \right), \nonumber \\
\ [ \hat{X}^4,\hat{X}^5 ] &=& \frac{i\beta^2}{4\alpha \ j'(j'+1)} \
\left( \frac{-i}{j(j+1)} \ \epsilon_{abc} \ \hat{X}^a \ \hat{X}^b \ 
\hat{X}^c \right)^{1/3}
\end{eqnarray}
Here the cubic root of a hermitian matrix is defined such that the resultant 
matrix is also hermitian. Although apparently there is no problem in the 
existence of the cubic root itself, more elaborate extension of 
(\ref{Ndimrepr}) may be needed. Nonetheless, the construction of the 
non-commutative product $\star$ in sec.3 can also be carried out for the  
configurations (\ref{extend}) straightforwardly. Moreover we expect that 
the corresponding values of integers in the $B$ field (\ref{BX}) are given by
$n_1= 2j'$ and $n_2=2j$. 

We also constructed a non-commutative product on $S^4$ which  corresponds  
to the Matrix configuration (\ref{Ndimrepr}). 
Because this configuration has the form of the tensor product of two 
matrices, we made these matrices to correspond to two $S^2$'s, and then 
directly constructed the 
non-commutative product $\star$ on $S^4$ in terms of that on $S^2$. 
We found that the manifold corresponding to the configuration 
(\ref{Ndimrepr}) is a twofold $S^4$. 

Here we should stress that to obtain a non-commutative product on $S^4$
we can neglect this doubling structure. The product (\ref{product1}) restricted to 
one of the doubled $S^4$ provides a full-fledged non-commutative product on 
$S^4$.  In this case, however,  the connection to the Matrix theory may be 
lost.
 
We also noticed that the product and 
the functions on $S^4$ have singularity (indeterninateness) on the circle
(\ref{circle}). The image of a point on this circle is $S^2$. In spite of this
singularity we can construct a non-commutative product and a finte action 
integral. We then worked out an explicit expression for the non-commutative 
product for the representation $j=1/2$.  

Let us now consider the commutative limit of the algebra of (\ref{extend}).
We expect that in the suitable large $j$, $j'$ limit the comutative geometry of
$S^4$ is recovered.  We will show this. When the constants $\alpha$, $\beta$ 
are related by $\beta = 2\alpha \sqrt{j(j+1)}$, the matrices $\hat{X}_0^A$ 
satisfy the constraint. 
\begin{equation}
(\hat{X}_0^A)^2= \hat{R}^2, \qquad \hat{R}=\alpha \ 
\sqrt{\frac{j(j+1)}{j'(j'+1)}}
\end{equation}
To make the radius $\hat{R}$ finite we will keep $\alpha$ and the ratio
$j/j'$ fixed. The commutators of $\hat{X}_0^A$ are given by
\begin{eqnarray}
\ [\hat{X}_0^a, \hat{X}_0^b] &=& i \frac{\alpha^2}{j^{'2}(j'+1)^2} \ 
\epsilon_{abc} \ (T_{(j')}^3)^2 \otimes T_{(j)}^c, \nonumber \\
\ [\hat{X}_0^a, \hat{X}_0^i] &=& i \frac{\alpha \beta}{2j^{'2}(j'+1)^2} \ 
\epsilon_{ij} \ T_{(j')}^{j-3} \otimes T_{(j)}^a, \nonumber \\
\ [\hat{X}_0^4, \hat{X}_0^5] &=& i \ \frac{\beta^2}{4j^{'2}(j'+1)^2} \ 
T_{(j')}^3 \otimes \bm{1}_{2j+1}
\end{eqnarray}
($a,b,c=1,2,3, \ \ i,j=4,5$) 
Since the matrix elements of $T_{(j)}^a$ and $T_{(j')}^a$ are at most order
${\cal O}(j)$ and ${\cal O}(j')$, respectively,  all the commutators vanish
like $1/j$ in the above mentioned $j, j' \rightarrow \infty$  limit. Therefore
this limit yields a commutative $S^4$ with a finite radius $\hat{R}$.

As for the next investigation  we are planning to study the 
following subjects. Our  Matrix configuration (\ref{Ndimrepr}) turned out to 
correspond to a twofold $S^4$.  We will study this structure in 
more details and seek for the possibility to construct new Matrix 
configurations which do not lead to a doubling structure.  
We are also planning to extend our construction of Matrix theory configuration 
to higher dimensional even sphere $S^{2m}$. 
Finally, we have not discussed a supersymmetric extension of 
our Matrix theory action (\ref{M_action2}). For this purpose we will need to 
introduce extra coordinates $X^6,\ldots, X^9$ in addition to fermions.

\section*{Acknowledgments}
\hspace{5mm}
The work of R.N. is supported in part by Grant-in-Aid (No.13135201) 
from the Ministry of Education, Science, Sports and Culture of Japan 
(Priority Area of Research (2)).


\section*{Appendix A: Multiplication Rule of functions on $(S^4)_{1/2}$}
\hspace*{5mm}
\begin{eqnarray}
 \frac{X^a}{R} \star \frac{X^b}{R}
  & = &
  \delta_{ab}+i\epsilon_{abc}\frac{X^c}{\Sq}, \qquad 
 \frac{X^i}{R} \star \frac{X^j}{R}
  = 
 \delta_{ij}+i\epsilon_{ij}\frac{\Sq}{R}, \label{mt1} \nonumber \\
 \frac{X^a}{R} \star \frac{X^i}{R}
 & = &
i\epsilon_{ij}\frac{X^j X^a}{R \Sq}, \qquad 
 \frac{X^i}{R} \star \frac{X^a}{R}
  = 
 -i\epsilon_{ij}\frac{X^j X^a}{R \Sq}, \nonumber \\
\frac{X^a}{\Sq} \star \frac{\Sq}{R}
& = &
\frac{X^a}{R}, \qquad \qquad \qquad 
\frac{\Sq}{R} \star \frac{X^a}{\Sq}
 = 
\frac{X^a}{R}, \nonumber \\
\frac{X^i}{R} \star \frac{\Sq}{R} 
& = &
-i \epsilon_{ij} \frac{X^j}{R}, \qquad \qquad 
\frac{\Sq}{R} \star \frac{X^i}{R}
 = 
i \epsilon_{ij} \frac{X^j}{R}, \nonumber \\
\frac{X^i}{R} \star \frac{X^j X^a}{R \Sq}
& = &
\delta_{ij}\frac{X^a}{\Sq}+i\epsilon_{ij}\frac{X^a}{R}, \nonumber \\
\frac{X^i X^a}{R \Sq} \star \frac{X^j}{R}
& = &
\delta_{ij}\frac{X^a}{\Sq}+i\epsilon_{ij}\frac{X^a}{R}, \nonumber \\
 \frac{X^a}{\Sq} \star \frac{X^b}{\Sq}
  & = &
  \delta_{ab}+i\epsilon_{abc}\frac{X^c}{\Sq}, \nonumber \\
\frac{X^i X^a}{R \Sq} \star \frac{X^b}{\Sq}
& = &
\delta_{ab}\frac{X^i}{R}+i\epsilon_{abc}\frac{X^i X^c}{R \Sq}, \nonumber \\
\frac{X^a}{\Sq} \star \frac{X^i X^b}{R \Sq}
& = &
\delta_{ab}\frac{X^i}{R}+i\epsilon_{abc}\frac{X^i X^c}{R \Sq}, \nonumber \\
\frac{X^a}{R} \star \frac{X^b}{\Sq}
& = &
\delta_{ab}\frac{\Sq}{R}+i\epsilon_{abc}\frac{X^c}{R}, \nonumber \\
\frac{X^a}{\Sq} \star \frac{X^b}{R}
& = &
\delta_{ab}\frac{\Sq}{R}+i\epsilon_{abc}\frac{X^c}{R}, \nonumber \\
\frac{X^i X^a}{R \Sq} \star \frac{\Sq}{R}
& = &
-i\epsilon_{ij}\frac{X^j X^a}{R \Sq}, \nonumber \\
\frac{\Sq}{R} \star \frac{X^i X^a}{R \Sq}
& = & 
i \epsilon_{ij}\frac{X^j X^a}{R \Sq}, \nonumber \\
\frac{X^a}{R} \star \frac{\Sq}{R}
& = &
\frac{X^a}{\Sq}, \qquad 
\frac{\Sq}{R} \star \frac{X^a}{R}
 = 
\frac{X^a}{\Sq}, \nonumber \\
\frac{X^a}{R} \star \frac{X^i X^b}{R \Sq}
& = &
i\epsilon_{ij}\delta_{ab}\frac{X^j}{R}
-\epsilon_{ij}\epsilon_{abc}\frac{X^j X^c}{R \Sq}, \nonumber \\
\frac{X^i X^a}{R \Sq} \star \frac{X^b}{R}
& = &
-i\epsilon_{ij}\delta_{ab}\frac{X^j}{R}
+\epsilon_{ij}\epsilon_{abc}\frac{X^j X^c}{R \Sq}, \nonumber \\
\frac{X^i X^a}{R \Sq} \star  \frac{X^j X^b}{R \Sq}
& = &
\delta_{ij}\delta_{ab}+i\delta_{ij}\epsilon_{abc}\frac{X^c}{\Sq} \nonumber \\
&& \qquad  +i\epsilon_{ij}\delta_{ab}\frac{\Sq}{R}
-\epsilon_{ij}\epsilon_{abc}\frac{X^c}{R}, \nonumber \\
\frac{\Sq}{R} \star \frac{\Sq}{R}
& = & 1
\label{mt23}
\end{eqnarray}
Here $\epsilon_{abc}$ and $\epsilon_{ij}$ are Levi-Civita symbols with
$\epsilon_{123}=\epsilon_{45}=+1$.

\section*{Appendix B: Coefficient Functions $L^{(n)}$}
\begin{eqnarray}
L^{(1)}_{a,b}
&=&
R^2 \delta_{ab} -X^a X^b +i R^2 \epsilon_{abc}\frac{X^c}{\Sq},
\label{L1}
\nonumber \\
L^{(2)}_{i,j}
&=&
R^2 \delta_{ij} -X^i X^j +i R^2 \epsilon_{ij}\frac{\Sq}{R},
\label{L2} \nonumber \\ 
L^{(3)}_{a,i}
&=&
-X^a X^i +i R^2 \epsilon_{ij}\frac{X^j X^a}{R\Sq},
\label{L3} \nonumber \\ 
L^{(4)}_{i,a}
&=&
-X^a X^i -i R^2 \epsilon_{ij}\frac{X^j X^a}{R\Sq},
\label{L4} \nonumber \\ 
L^{(5)}_{ab,c}
&=&
\frac{R^2-\Sq^2}{2}(X^a\delta_{bc}+X^b\delta_{ac})
-\frac{R^2-\Sq^2}{\Sq^2}X^a X^b X^c \nonumber\nonumber \\&&
+i\frac{R^2-\Sq^2}{2}(X^b \epsilon_{acd}+X^a\epsilon_{bcd})\frac{X^d}{\Sq},
\nonumber \\ 
L^{(6)}_{a,bc}
&=&
\frac{R^2-\Sq^2}{2}(X^c\delta_{ab}+X^b\delta_{ac})
-\frac{R^2-\Sq^2}{\Sq^2}X^a X^b X^c \nonumber \nonumber \\&&
+i\frac{R^2-\Sq^2}{2}(X^c \epsilon_{abd}+X^b\epsilon_{acd})\frac{X^d}{\Sq},
 \nonumber \\ 
L^{(7)}_{ab,i}
&=&
-\frac{1}{2}X^a X^b X^i +i \frac{R}{\Sq}\epsilon_{ij}X^j X^a X^b,
\nonumber \\ 
L^{(8)}_{a,bi}
&=&
-\Sq^2 X^i\delta_{ab} +X^a X^b X^i -i \Sq \epsilon_{abc}X^c X^i
+i R\Sq \delta_{ab}\epsilon_{ij}X^j \nonumber \nonumber \\&&
-i\frac{R}{\Sq}\epsilon_{ij}X^j X^a X^b
-R \epsilon_{ij}\epsilon_{abc}X^j X^c,
\nonumber \\\ 
L^{(9)}_{ai,b}
&=&
-\Sq^2 X^i\delta_{ab} +X^a X^b X^i -i \Sq \epsilon_{abc}X^c X^i
-i R\Sq \delta_{ab}\epsilon_{ij}X^j  \nonumber \nonumber \\&&
+i\frac{R}{\Sq}\epsilon_{ij}X^j X^a X^b
+R \epsilon_{ij}\epsilon_{abc}X^j X^c,
\nonumber \\ 
L^{(10)}_{i,ab}
&=&
-\frac{1}{2}X^a X^b X^i - i \frac{R}{\Sq}\epsilon_{ij}X^j X^a X^b,
 \nonumber \\ 
L^{(11)}_{ai,j}
&=&
\frac{\Sq^2}{2}X^a \delta_{ij}+i R^2 X^a \epsilon_{ij}\frac{\Sq}{R},
 \nonumber \\ 
L^{(12)}_{a,ij}
&=&
-i\frac{R}{2}X^a (\epsilon_{ik}X^j+\epsilon_{jk}X^i)\frac{X^k}{\Sq},
 \nonumber \\ 
L^{(13)}_{i,aj}
&=&
\frac{\Sq^2}{2}X^a \delta_{ij}+i R^2 X^a \epsilon_{ij}\frac{\Sq}{R},
 \nonumber \\ 
L^{(14)}_{ij,a}
&=&
i\frac{R}{2}X^a (\epsilon_{ik}X^j+\epsilon_{jk}X^i)\frac{X^k}{\Sq},
 \nonumber \\ 
L^{(15)}_{ij,k}
&=&
-i \frac{R\Sq}{2}(\delta_{ik}\epsilon_{jl}+\delta_{jk}\epsilon_{il})X^l
+\frac{\Sq^2}{4}(\epsilon_{ik}\epsilon_{jl}+\epsilon_{il}\epsilon_{jk})X^l,
\nonumber \\ 
L^{(16)}_{i,jk}
&=&
i \frac{R\Sq}{2}(\delta_{ij}\epsilon_{kl}+\delta_{ik}\epsilon_{jl})X^l 
-\frac{\Sq^2}{4}(\epsilon_{ij}\epsilon_{kl}+\epsilon_{ik}\epsilon_{jl})X^l,
 \nonumber \\ 
L^{(17)}_{ab,cd}
&=&
\frac{\Sq^2}{4}(R^2-\Sq^2)(\delta_{ac}\delta_{bd}+\delta_{ad}\delta_{bc})
+\frac{R^2-\Sq^2}{2\Sq^2}X^a X^b X^c X^d \nonumber \nonumber \\&&
+i\frac{R^2-\Sq^2}{4}(X^a X^c \epsilon_{bde}+X^a X^d\epsilon_{bce}
+X^b X^d \epsilon_{ace}+X^b X^c\epsilon_{ade})\frac{X^e}{\Sq} 
\nonumber \nonumber \\&&
-\frac{R^2-\Sq^2}{4}(\epsilon_{ace}\epsilon_{bdf}
+\epsilon_{ade}\epsilon_{bcf})X^e X^f,
 \nonumber \\ 
L^{(18)}_{ij,kl}
&=&
-\frac{\Sq^4}{4}(\delta_{ik}\delta_{jl}+\delta_{il}\delta_{jk})
+\frac{\Sq^4}{4}(\epsilon_{ik}\epsilon_{jl}+\epsilon_{il}\epsilon_{jk}),
 \nonumber \\ 
L^{(19)}_{ab,ci}
&=&
-\frac{\Sq^2}{2}X^i(X^b\delta_{ac}+X^a\delta_{bc})
-i\frac{\Sq}{2}X^i(X^b\epsilon_{acd}+X^a\epsilon_{bcd})X^d\nonumber \\&&
+i\frac{R\Sq}{2}(X^b\delta_{ac}+X^a\delta_{bc})\epsilon_{ij}X^j
-\frac{R}{2}(X^b\epsilon_{acd}+X^a\epsilon_{bcd})\epsilon_{ij}X^j X^d,
\nonumber \nonumber \\ 
L^{(20)}_{ai,bc}
&=&
-\frac{\Sq^2}{2}X^i(X^c\delta_{ab}+X^b\delta_{ac})
-i\frac{\Sq}{2}X^i(X^c\epsilon_{abd}+X^b\epsilon_{acd})X^d \nonumber \nonumber \\&&
-i\frac{R\Sq}{2}(X^c\delta_{ab}+X^b\delta_{ac})\epsilon_{ij}X^j
+\frac{R}{2}(X^c\epsilon_{abd}+X^b\epsilon_{acd})\epsilon_{ij}X^j X^d,
 \nonumber \\ 
L^{(21)}_{ab,ij}
&=&
-\frac{1}{2}\epsilon_{ik}\epsilon_{jl}X^k X^l X^a X^b,
\nonumber \\ 
L^{(22)}_{ij,ab}
&=&
-\frac{1}{2}\epsilon_{ik}\epsilon_{jl}X^k X^l X^a X^b,
 \nonumber \\ 
L^{(23)}_{ai,bj}
&=&
R^2\Sq^2\delta_{ij}\delta_{ab}-\Sq^2 X^i X^j\delta_{ab}
-\frac{R^2-\Sq^2}{2}X^a X^b \delta_{ij} \nonumber \nonumber \\&&
+i R^2\Sq\delta_{ij}\epsilon_{abc}X^c
-i \Sq X^i X^j \epsilon_{abc}X^c \nonumber \nonumber \\&&
+iR^2\Sq^2\delta_{ab}\epsilon_{ij}\frac{\Sq}{R}
-R\Sq^2\varepsilon_{ij}\epsilon_{abc}X^c,
 \nonumber \\ 
L^{(24)}_{ai,jk}
&=&
\frac{\Sq^2}{4}X^a(X^k\delta_{ij}+X^j\delta_{ik}),
\nonumber \\ 
L^{(25)}_{ij,ak}
&=&
\frac{\Sq^2}{4}X^a(X^j\delta_{ik}+X^i\delta_{jk})
\end{eqnarray}

\newpage



\begin{thebibliography}{99}
\bibitem{CLT} J.~Castelino, S.~Lee and W.~Taylor IV, {\it Londitudinal 
5-branes as 4-spheres in Matrix theory}, Nucl.Phys.{\bf B526} (1998) 334-350,
hep-th/9712105.
\bibitem{BFSS} T.~Banks, W.~Fischler, S.~H.~Shenker and L.~Suskind, 
{\it M theory as a matrix model: a conjecture}, Phys.Rev. {\bf D55} (1997) 
5112-5128, hep-th/9610043.
\bibitem{IIB} N.~Ishibashi, H.~Kawai, Y.~Kitazawa and A.~Tsuchiya, 
{\it A Large N reduced model as superstring}, Nucl. Phys. {\bf B498} (1997) 
467-491, hep-th/9612115. 
\bibitem{GKP} H.~Grosse, C.~Klim\v{c}$\acute{\imath}$k and P.~Pre\v{s}najder,
{\it On Finite 4D Quantum Field Theory in Non-Commutative Geometry}, 
Commun.Math.Phys. {\bf 180} (1996) 429-438, hep-th/9602115.
\bibitem{Oconner} J.~Medina and D.~O'Connor, {\it Scalar field Theory on fuzzy $S^4$}, JHEP {\bf 0311} (2003) 051, hep-th/0212170.
\bibitem{HoRangoolam} P.~M.~Ho and S.~Rangoolam, {\it Higher dimensional geometries from matrix brane constructions}, Nucl.Phys. {\bf B627} (2002) 266-288, 
hep-th/0111278. 
\bibitem{Rangoolam} S.~Rangoolam, {\it On spherical harmonics for fuzzy 
spheres in diverse dimensions}, Nucl.Phys. {\bf B610} (2001) 461-488, 
hep-th/0105006;{\it Higher dimensional geometries related Fuzzy 
odd-dimensional sheres}, JHEP {\bf 0210} (2002) 064, hep-th/0207111.
\bibitem{CNYang} C.~N.~Yang, {\it Generalization of Dirac's monopole to $SU_2$
gauge fields}, J.Math.Phys. {\bf 19} (1978) 320.
\bibitem{ZH} S.~-C.~Zhang and J.~P.~Hu, {\it A four-dimensional generalization
of the quantum Hall effect}, Science {\bf 294} (2001) 823.
\bibitem{Madore} J.~Madore, Class. Quantum Grav. {\bf 9} (1992) 69.
\bibitem{Presnejder} P.~Pre\v{s}najder, {\it The origin of chiral anomaly and
the noncommutative geometry}, 
J.Math.Phys. {\bf 41} (2000) 2789-2804, hep-th/9912050.
\bibitem{HNT} K.~Hayasaka, R.~Nakayama and Y.~Takaya, {\it A new 
noncommutative product on the fuzzy two-sphere corresponding to the unitary 
representation of $SU(2)$ and the Seiberg-Witten map}, Phys. Lett. {\bf B553} 
(2003)109-118, hep-th/0209240.
\bibitem{BMN} D.~Berenstein, J.~Maldacena and H.~Nastase, {\it Strings in flat
space and pp waves from ${\cal N}=4$ Super Yang Mills}, 
JHEP {\bf 0204} (2002) 013, hep-th/0202021.
\bibitem{Meyers} R.~C.~Meyers, {\it Dielectric Branes}, JHEP {\bf 9912} 
(1999) 022,
hep-th/9910053.
\bibitem{SW} N.~Seiberg and E.~Witten, {\it String Theory and Noncommutative 
Geometry}, JHEP {\bf 9909} (1999) 032, hep-th/9908142. 
\bibitem{Kontsevich} M.~Kontsevich, {\it Deformation quantization of Poisson 
manifolds}, q-alg/9709040.
\end{thebibliography}
\end{document}